\documentstyle[editedvolume,psfig]{crckapb10}

\newcommand{\eg}{{\it e.g.\ }}
\newcommand{\degr}{\mbox{$^\circ$}}

\newcommand{\kms}{\mbox{\,km\,s$^{-1}$}}

\newcommand{\cm}{\rm\,cm}

\newcommand{\kmsMpc}{{\rm\,km\,s$^{-1}\,$Mpc$^{-1}$}}

\newcommand{\kpc}{{\rm\,kpc}}

\newcommand{\OII}{\mbox{[O\,{\sc ii}]$\lambda$3727}}
\newcommand{\Oii}{\mbox{[O\,{\sc ii}]}}

\newcommand{\farcs}{\hbox{$.\!\!^{\prime\prime}$}}

\newcommand{\fasec}{\,\farcs}

\newcommand{\C}[1]{3C\,#1}
\newcommand{\FR}[1]{FR\,II\ }

\def\simlt{\mathrel{\mathchoice {\vcenter{\offinterlineskip\halign{\hfil
$\displaystyle##$\hfil\cr<\cr\sim\cr}}}
{\vcenter{\offinterlineskip\halign{\hfil$\textstyle##$\hfil\cr<\cr\sim\cr}}}
{\vcenter{\offinterlineskip\halign{\hfil$\scriptstyle##$\hfil\cr<\cr\sim\cr}}}
{\vcenter{\offinterlineskip\halign{\hfil$\scriptscriptstyle##$\hfil\cr<\cr\sim\cr}}}}}
\def\simgt{\mathrel{\mathchoice {\vcenter{\offinterlineskip\halign{\hfil
$\displaystyle##$\hfil\cr>\cr\sim\cr}}}
{\vcenter{\offinterlineskip\halign{\hfil$\textstyle##$\hfil\cr>\cr\sim\cr}}}
{\vcenter{\offinterlineskip\halign{\hfil$\scriptstyle##$\hfil\cr>\cr\sim\cr}}}
{\vcenter{\offinterlineskip\halign{\hfil$\scriptscriptstyle##$\hfil\cr>\cr\sim\cr}}}}}

\newcommand{\apj}{{\it Astrophys. J.}}
\newcommand{\apjs}{{\it Astrophys. J. Suppl.}}
\newcommand{\apjl}{{\it Astrophys. J. Lett.}}
\newcommand{\aj}{{\it Astron. J.}}
\newcommand{\mn}{{\it Mon. Not. R. Astron. Soc.}}
\newcommand{\aap}{{\it Astron. Astrophys.}}
\newcommand{\nat}{{\it Nature}}

\psfull

\begin{opening}
\title{Probing the Warm Gas in \protect\\
High Redshift Radio Galaxies} \protect\\
\subtitle{New Results on the Alignment Effect}

\author{Mark J. Neeser}
\author{Hans Hippelein}
\author{Klaus Meisenheimer}
\institute{Max-Planck-Institut f\"ur Astronomie \\
K\"onigstuhl 17, 69117 Heidelberg, Germany}

\end{opening}

\runningtitle{The Warm Gas in High Redshift Radio Galaxies}

\begin{document}

\noindent{\bf Abstract.}
We report on the results of an in depth investigation of the extended
\OII\ line-emission in 11 3CR radio galaxies (0.5 $<$ z $<$ 1.1).
Using a Fabry-Perot etalon to obtain both the kinematics and
morphology of the \Oii\ gas, our goal was to find the 
mechanisms responsible for the creation and excitation of 
this warm gas, and the source of its alignment with the 
radio emission.

\section{Introduction}


Powerful radio galaxies are often associated with complex extended emission-line
regions that can have linear sizes of up to several hundred kpc.  
The first systematic imaging surveys by Baum et al. (1988) and
McCarthy et al. (1987) (low and high z 3CR sources, respectively), and by
Chambers et al. (1987) (4C sources) not only revealed a close relationship
between the radio power and the optical emission-line luminosity, but also
a tendency for the extended gas to share the same axis as that of its double-lobed
radio source.  Furthermore, the fraction of radio galaxies displaying this
emission-line {\it alignment effect} rapidly changes from a few at low redshifts, 
to nearly all for z\,$>$\,0.3.

The combination of the novelty of this discovery with the often
spectacular morphologies of extended emission-line regions, and their potential
effects on the evolution and formation of radio galaxies, attracted lively
debate, little consensus, and a large number of possible explanations for this
phenomenon.
However, since the line-emission regions in high redshift radio
galaxies, by virtue of their large intrinsic luminosities and large spatial
extents, provide one of the best methods for probing the warm gas at
early epochs,
the importance of
understanding the alignment effect cannot be understated.
Since this phenomena arises in extended, highly dynamic gas, an
investigation that combines well-resolved morphologies with kinematics would
go far in shedding new light on the question of the origin of the warm
gas, its source of excitation, its influence on the host galaxy, and the cause 
of the emission-line/radio source alignment.

\section{Observational Methods}

Selecting the most extended emission-line sources with z\,$\simgt$\,0.5 from
McCarthy et al. (1995), we imaged a subsample of 3CR radio sources (see table 1)
using a Fabry-Perot (FP) interferometer with a spectral
and spatial resolution of $\sim$400 kms$^{-1}$ and $\simlt$1\fasec 6, respectively.
By stepping the FP along the \OII\ emission-line (typically 8--10 wavelength 
settings across the line), we were able to simultaneously map the velocity 
field and morphology of the ionized gas and create a representative sample of 
11 radio galaxies observed with unprecedented detail.  To investigate the
continuum morphologies, as well as subtract their contribution to the line-emission
images, intermediate-band ($\lambda/\Delta\lambda\simeq$40) line-free exposures
were also obtained on either side of the redshifted \OII\ line.

\vspace*{-0.5cm}
\begin{table}[h]
\begin{center}
\caption{\small Fabry-Perot \Oii\ Sample}
\begin{tabular}{lcc}
\hline
Source  &Redshift  &Size of \Oii\ Region  \\
\hline
3C\,34     &   0.69  &  17{\tt "} \\
3C\,44     &   0.66  &  10{\tt "} \\
3C\,54     &   0.83  &  5{\tt "} \\
3C\,124    &   1.08  &  4{\tt "} \\
3C\,169.1  &   0.63  &  8{\tt "} \\
3C\,265    &   0.81  &  35{\tt "} \\
3C\,337    &   0.64  &  13{\tt "} \\
3C\,352    &   0.81  &  12{\tt "} \\
3C\,368    &   1.13  &   9{\tt "} \\
3C\,435A   &   0.47  &  16{\tt "} \\
3C\,441    &   0.71  &  5{\tt "} \\
\hline
\end{tabular}
\end{center}
\label{tab:sample}
\end{table}
\vspace*{-0.7cm}

\section{Main Results From the Entire \Oii\ Sample}

Somewhat surprisingly, up
to 40\% of the radio sources investigated have line-emission regions
dominated by the effects of galaxy--galaxy interactions.  For these objects
we are able to show that the popularly accepted models for exciting the
line-emission regions are inadequate to explain all of their observed \Oii\
features.  Instead, we propose a
new model in which a close, strong interaction with a companion galaxy
exchanges material with the central radio galaxy, creating a complex morphology
of bridges, tails and extended knots along the interaction axis.
The passage of the companion through the halo of
the host source shock heats the supplied gas
and the ambient medium, thus creating parts or all of the observed
emission-line region and its complex
velocity structure.  The interaction itself may also be responsible for
triggering the radio source of the central galaxy.
 
Since this new model applies to a significant fraction of the radio source
sample, it is important to explain how galaxy--galaxy interactions (an
intrinsically geometrically random phenomena) can give rise to alignments
between the radio emission and the warm gas.
It is plausible that galaxies undergoing gravitational interactions, when aligned
with their radio sources, will be preferentially selected by the
flux-limits of the 3CR catalogue.  Most models of double radio sources predict
that the radio luminosity of a source will be increased if it expands into a
denser gaseous medium (see Eales 1992 for a discussion of this effect).  
Therefore, since only radio galaxies in which one lobe
lies near to the line-emission gas supplied by the companion galaxy will
experience an enhancement of its radio emission, flux-limited samples will
preferentially contain aligned interaction galaxies.  The fact that our sources
consistently have their
brighter and/or closer radio lobe on the same side of the central source as
most of its line-emission gas, supports this scenario.  Another important clue
lies in the fact that all of the interaction galaxies are near to the flux-limit
of the 3CR catalogue.  This implies that without the
radio brightness asymmetries these objects would not have been detected by
this survey.  By relating the typical radio luminosity and lobe size asymmetries
to density differences, we are also able to show that only modest density 
contrasts ($\frac{\rho_{\rm [OII] gas}}{\rho_{\rm IGM}} \simlt 7$),
between a radio lobe expanding near to the line-emission gas and its counter 
part expanding into the ambient intergalactic medium on the other side of the 
central galaxy, are necessary.

For the remaining radio galaxies
variations of the more conventional mechanisms---photoionizing
radiation from a central, hidden active nucleus, or a direct interaction
between the radio source and the ambient medium---are used to explain the
excitation and alignment of the \OII\ emitting gas.
Since each of these three models results
in ionized gas morphologies and kinematics that are unique, we are able to use
these characteristics to define three distinct classes of line-emission galaxy.
The archetype source in each class is shown in bold type.

\begin{itemize}
\item[1.] Strong galaxy--galaxy interactions.  The sources whose
line-emission regions are dominated by this model show
one-sided line-emission morphologies distinguished by multiple
components, \OII\ bridges, and extended linear features.
A large range of line-emission sizes and degree of alignment, as well as complex
velocity  structures are among the features typical of these objects.
The intrinsically one-sided nature of the interaction model provides a natural
explanation for the very large line-emission brightness, morphology, and
velocity asymmetries that define this class (\eg\ {\bf \C{169.1}}, \C{435A}, 
\C{44}, and the inner region of \C{265}).
\item[2.] Photoionization by a central active galaxy, hidden from our view
by an obscuring torus.
These objects are among the largest and most symmetric sources
and are characterized by conical, or bi-conical line-emission structures, and
relatively quiescent line widths (\eg\ {\bf \C{34}} and \C{265}).
\item[3.] A direct shock interaction between the radio source and the ambient
gaseous medium.  Galaxies dominated by this model
tend to be very well-aligned and have the greatest degree of symmetry in their
line-emission luminosity, morphology, and kinematics.  They are also
characterized by having the largest line widths of all of our sample sources.
For a number of galaxies we have also proposed a weaker version of this
model, that involves an interaction between the lateral expansion/backflow of
the radio lobes and the \Oii\ gas (\eg\ {\bf \C{368}}, \C{352}, \C{34}, and 
\C{337}).


\end{itemize}

The characteristics of each class are sufficiently unique that they can be 
used to match
a radio galaxy with the mechanisms responsible for the excitation and alignment
of its line-emission regions, despite the fact that the line-emission regions of 
some of our radio galaxies are sufficiently complex that multiple 
mechanisms for forming and exciting the warm gas are necessary.  This fact,
together with the three mechanisms needed to explain all of our observations, are
a warning against a single, universal model for the alignment
effect in all powerful radio galaxies.

Noticeably absent from the list of mechanisms for explaining the
emission-line excitation is the popular jet-induced starburst scenario.  
In this model the radio source propagates through the ambient medium, compresses
the gas through its bowshock or overpressure cocoon, and triggers a
burst of star formation which is able to {\it in situ} photoionize the underlying
\Oii\ gas (\eg\ McCarthy et al. 1987; DeYoung 1989;
Begelman \& Cioffi 1989).  Using the spectral synthesis models of Bruzual \&
Charlot (1993),
we have computed the spectral energy distribution of a starburst constructed 
to maximize its output of ionizing radiation.  Our observed \OII\ flux constrains
the mass of this burst, allowing a prediction of its continuum 
signature.  Although this computed continuum flux should be easily detectable
in our broad-band images, the fact that we find little or no continuum
underlying the line-emission components, strongly argues against this model.

\section{An Example of a High Redshift Photoionization Cone}

\begin{figure*}[h]
\centerline{
\psfig{figure=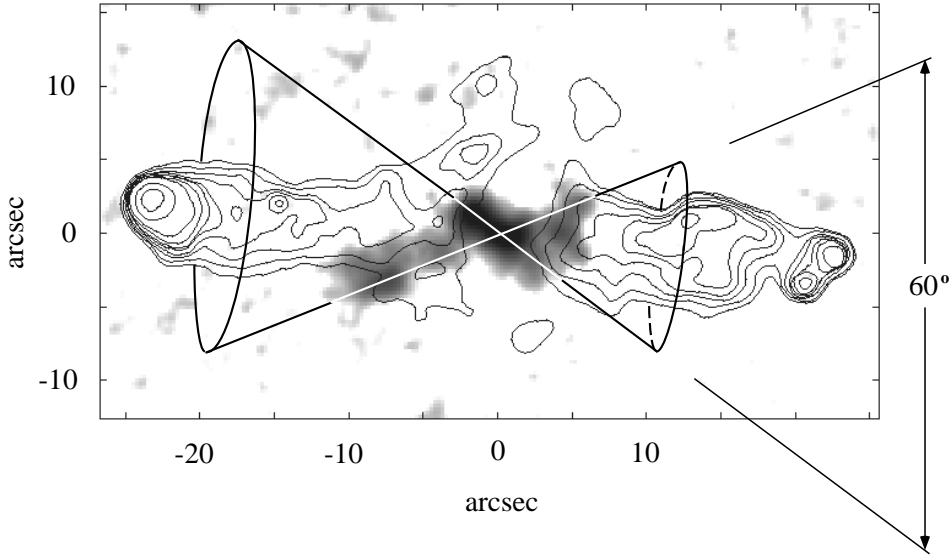,width=13.2cm}}
\vspace*{-0.5cm}
\caption[\C{34}: Photoionization Cone]
{Our proposed photoionization cone on the grayscale \Oii\
image of \C{34}.  The 20\,cm radio map of Neff et al. (1995) is shown as
contours.  The apex of the cone is located at the position of the central 
continuum source (the location of the hidden AGN).  An opening angle of
60\degr\ is the minimum required to photoionize the observed line-emission.
\vspace*{-0.2cm}
\label{fig:3c34cone}}

\end{figure*}

To give a specific example from the sources discussed in the previous section,
we will give a brief description of our results for \C{34} (z=0.689).  A more
thorough presentation of this galaxy is given in Neeser et al. (1997).
The excitation of the 120\,kpc-sized (H$_{\circ}$=50\kmsMpc, q$_{\circ}$=0.5)
line-emission region of this radio galaxy, is the result of photoionizing
radiation emanating anisotropically from a hidden AGN (class 2 emission-line
region).  This is indicated by the distinctively bi-conical morphology of
\C{34}'s \Oii\ gas.  By placing the apex of a symmetrical bi-cone at the 
position of the central continuum source, we find that we can
connect 5 distinct line-emission knots/extensions on both sides of the source,
and symmetrically straddle the radio source axis (see figure \ref{fig:3c34cone}).

This model can also explain a number of smaller scale line-emission features:
intensity profiles of components that are steepened toward the central source,
and components that show various degrees of `shadowing'
outward from the central ionizing source, indicating that the directed UV emission
is radiation bounded in parts of the \Oii\ region.
%
%
A simple photoionization model
shows that this interpretation is energetically viable on these
length scales, when the cumulative covering factors of the outermost line-emission
components approach unity.
The luminosity of the hidden central AGN, necessary to account for the
observed \OII\ luminosity, is compatible with that of a typical 3CR quasar
at a similar redshift.  

\begin{figure*}[h]
\centerline{
\psfig{figure=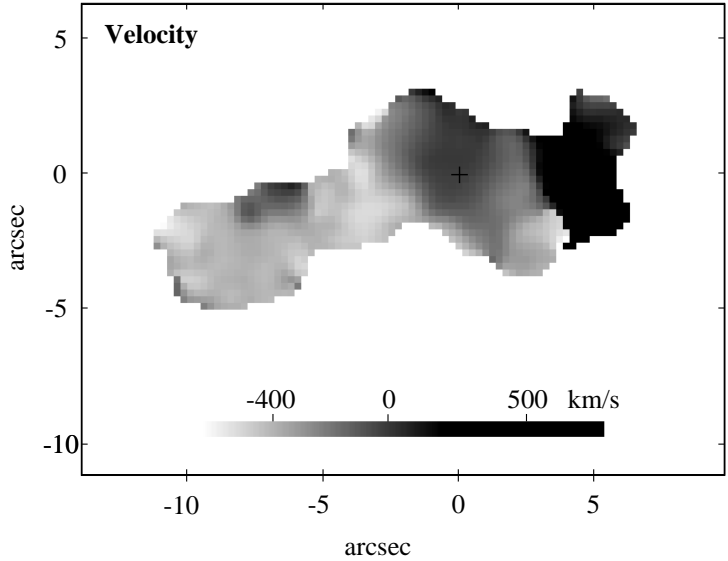,width=11.7cm}}
\vspace*{-0.5cm}
\caption[\C{34}: Kinematics]{A grayscale representation of the radial velocities
of the \OII\ line-emission in \C{34}.  The kinematical and positional origin is 
defined by the optical continuum counterpart of the radio
galaxy (shown with a cross). At the bottom of the image a
scale is given matching the diagram's grayscale to velocities in \kms.
\vspace*{-0.3cm}
\label{fig:3c34velwidt}}
\end{figure*}

Although this interpretation can account for the
excitation and parts of the warm gas morphology, it is insufficient to explain the
observed velocity and line-width structures.  The simplest photoionization
model would assume that the ionization cone is merely illuminating the random
velocity gas clumps that exist in a typical cluster environment.  The difficulties
with this interpretation are that the line-emission in \C{34} is
not distributed in discrete clumps, and that the
eastern component shows a remarkably flat velocity structure across a length
of more than 70\kpc\ (see figure \ref{fig:3c34velwidt}).
%
%
%
%
This indicates that we require a {\em single} mechanism to act on a 
large fraction of gas simultaneously.  We therefore propose that the radio 
source has, through the bulk motions of its lateral expansion and backflow, 
enmasse swept-up the gas that existed in the environment of \C{34}.  In this 
way the gas of the eastern \Oii\ line-emission region has been compressed,
pushed to the outer edge of the radio lobes, and given a bulk velocity
that is constant across the entire region.
A close correlation between the line-emitting gas and the outer edges of the
20\cm\ radio emission, as well as the radio depolarization associated with
the \Oii\ gas \cite{Johnson95}, also support this interpretation.

In our photoionization scenario there exists a direct
cause and effect relationship between the radio source and the ionization cone
that leads to their alignment in \C{34}.
It is possible to imagine that the nucleus was initially surrounded
by a cloud opaque to ionizing radiation in all directions.
When the radio jet turned on it plowed through the cloud and opened up a
low density channel.  As the radio lobes grew in size the increased density
of the swept-up gas allows it to effectively absorb
the incident ionizing radiation from the central AGN; this, in turn, can then
effectively escape along the cleared out, low density channel created by the
radio source.

The obvious conical structure in \C{34}, though previously unobserved in
high redshift, powerful radio galaxies, is well-known in 11 low redshift
Seyferts (\eg\ Wilson \& Tsvetanov 1994).  The tight alignment between the cone 
and radio axes found in these sources ($\Delta$PA$_{\rm mean}$=6\degr;
Wilson \& Tsvetanov 1994) is also true for \C{34}.  A fundamental difference, 
however,
is that the Seyfert ionization cones show line-emission gas across the entire
lateral extent of their opening angles.  The fact that \C{34}'s radio power
is more than four orders of magnitude greater and hence capable of effectively
sweeping out the IGM of the radio galaxy, and confining the line-emitting gas
to its edges, is a plausible explanation for this difference.

\end{document}